# $^{56}$Ni, Explosive Nucleosynthesis, and SNe Ia Diversity

**James W Truran[1,2], Ami S Glasner[3], and Yeunjin Kim[1]**

[1]Department of Astronomy and Astrophysics, University of Chicago, 5640 South Ellis Avenue, Chicago, IL 60637, USA

[2]Physics Division, Argonne National Laboratory, Argonne, IL 60439, USA

[3]Racah Institute of Physics, The Hebrew University, Jerusalem 91904, ISRAEL

E-mail: truran@nova.uchicago.edu

**Abstract**. The origin of the iron-group elements titanium to zinc in nature is understood to occur under explosive burning conditions in both Type Ia (thermonuclear) and Type II (core collapse) supernovae. In these dynamic environments, the most abundant product is found to be $^{56}$Ni (($\tau$ = 8.5 days) that decays through $^{56}$Co ($\tau$ = 111.5 days) to $^{56}$Fe. For the case of SNe Ia, the peak luminosities are proportional to the mass ejected in the form of $^{56}$Ni. It follows that the diversity of SNe Ia reflected in the range of peak luminosity provides a direct measure of the mass of $^{56}$Ni ejected. In this contribution, we identify and briefly discuss the factors that can influence the $^{56}$Ni mass and use both observations and theory to quantify their impact. We address specifically the variations in different stellar populations and possible distinctions with respect to SNe Ia progenitors.

## 1. Introduction

Dating back to the early work by Hoyle and Fowler [1], the 'standard' model for a Type Ia supernova has been assumed to involve the explosion of a carbon-oxygen (CO) white dwarf, with iron peak nuclei being a major product. The history of our understanding of supernova iron production began with Hoyle's [2] identification of the nucleosynthesis mechanism with a "nuclear statistical equilibrium" process occurring at high temperatures and densities in stellar cores. The classic papers by Burbidge et al. [3] and Cameron [4] both expanded upon this picture and sought to identify the physical conditions (T, $\rho$, $Y_e$) that provided the best fit to Solar System abundances for an iron peak dominated by $^{56}$Fe. These efforts encountered difficulties. Subsequently Cameron [5] noted the critical dependence of explosive nucleosynthesis of iron-peak nuclei on neutron enrichment and Truran, Arnett and Cameron [6] demonstrated that representative supernova conditions yield $^{56}$Ni as the principal product *in situ*. The first paper to incorporate the effects of $^{56}$Ni decay in a study of SNe Ia light curves was that of Colgate and McKee [7].

Why $^{56}$Ni? The physics that dictates the *in situ* formation of a $^{56}$Ni dominated iron abundance peak under explosive supernova conditions is straightforward. For both Type Ia and Type II supernovae, the pre-explosion compositions involve primarily nuclei of Z=N, viz: $^{12}$C, $^{16}$O, and $^{28}$Si. Explosive burning at temperatures T > 4 x $10^9$ K typically occurs on timescales of fractions of seconds, on which timescales weak interaction processes effecting 'neutronization' of the matter proceed too slowly to convert any significant fraction of protons to neutrons. It follows that the main products of explosive burning *in situ* are proton-rich nuclei of Z = N, viz: $^{44}$Ti, $^{48}$Cr, $^{52}$Fe, $^{56}$Ni, $^{60}$Zn, and $^{64}$Ge. An intriguing outgrowth of these same nucleosynthesis conditions is written in the isotopic compositions of iron peak elements. Specifically, in the relevant supernova environments $^{50}$Cr is formed directly as $^{50}$Cr, while $^{52}$Cr and $^{53}$Cr are decay products of $^{52}$Fe and $^{53}$Fe, $^{54}$Fe is synthesized as $^{54}$Fe, while $^{56}$Fe and $^{57}$Fe are decay products of $^{56}$Ni and $^{57}$Ni, and $^{58}$Ni is synthesized as $^{58}$Ni, while $^{60}$Ni, $^{61}$Ni, and $^{62}$Ni are decay products of $^{60}$Zn, $^{61}$Zn, and $^{62}$Zn – but the resulting isotopic patterns for these respective elements are consistent with the isotopic composition of Solar System/Cosmic matter. This characteristic of the isotopic patterns of iron-peak nuclei formed in explosive nucleosynthesis is apparent in the figure below that identifies both the proton-rich nuclei along the alpha line and the stable isotopes in the iron peak regime to which they decay. This behavior strongly confirms that the iron-peak elements are indeed formed under neutron-poor ($Y_e \approx 0.5$) explosive burning conditions, for which $^{56}$Ni is the dominant product.

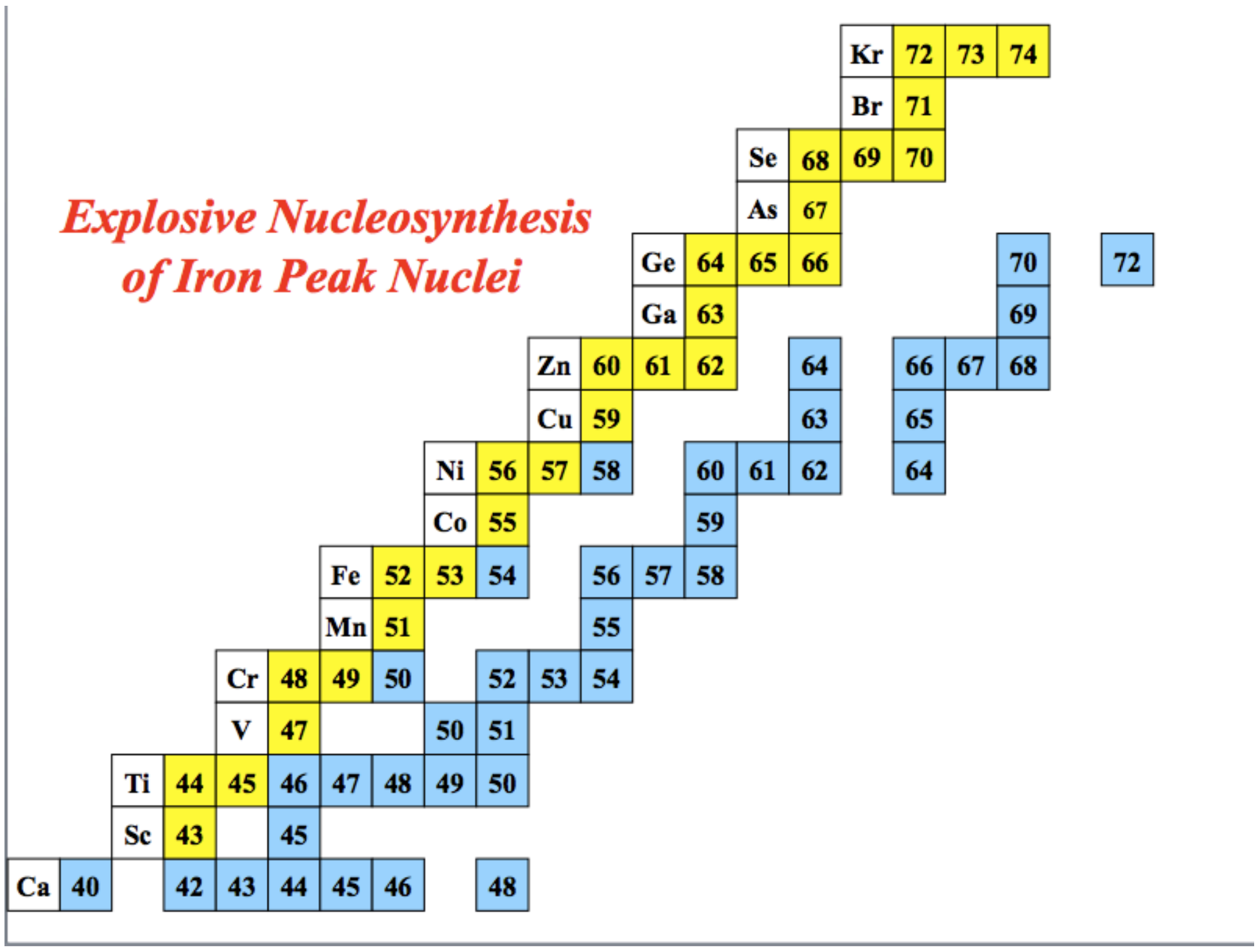

Figure 1: The nuclear chart in the iron peak regime from calcium to krypton. Stable isotopes are shown in blue. Unstable isotopes in proximity to the Z=N line that constitute significant radiogenic decay parents of abundant stable isotopes of iron-peak nuclei are shown in yellow. This illustrates the character of the synthesis of iron-peak nuclei occurring under proton-poor burning conditions in explosive supernova environments.

In this paper we are particularly concerned with Type Ia (thermonuclear) supernovae. These brightest of supernovae can be used effectively as distance indicators. Renewed interest in these events was aroused by the fact that their use over the past decade has provided compelling evidence for an accelerating cosmic expansion (Reiss et al. [8], Perlmutter at al. [9]). In this context, the Phillips [10] relation allows one to compensate for the observed variations (e.g. "diversity") in peak supernova luminosity (brighter SNe Ia decline more slowly) to provide a "standard candle." The critical factor here again is the synthesis of iron ($^{56}$Fe) as $^{56}$Ni, with the peak luminosity being proportional to the mass of $^{56}$Ni ejected (Arnett [11]).

## 2. The Observed Diversity of Type Ia Supernovae

One of the most striking characteristics of the observed diversity of Type Ia supernovae is the broad range in their luminosities at maximum. This diversity in peak luminosity and its correlation with the properties/characteristics of the astronomical environments in which SNe Ia occur has now been well documented observationally. We now fully appreciate that, while Type Ia supernovae are the only type of supernova event seen to occur in early type (elliptical) galaxies, they are also observed to occur in younger stellar populations. The view that SNe Ia can have short-lived progenitors (~ $10^8$ years) was first emphasized many years ago by Oemler and Tinsley [12], based upon their recognition that the SNe Ia rate per unit mass is very high in irregular galaxies and roughly proportional to their present star formation rates in spiral galaxies. More recently, Scannapieco and Bildsten [13] have essentially expanded upon this in their discussion of a two component - prompt and extended – Type Ia supernova rate.

The observational support for both short lived and long lived progenitor models for SNe Ia has been noted and explored by a number of sources [14,15,16,17]. The observed trends relevant to our discussions in this paper can be briefly summarized as follows: (1) A rather broad range of peak Type Ia supernova luminosities is observed quite independent of galaxy type (environment), that is both for young stellar populations (e.g. spiral and irregular galaxies) and for older stellar populations (e.g. elliptical or S0 galaxies). For spiral galaxies, the spread can be greater than a factor ~ 5 (stretch factors ~ 0.7 to 1.15), while for elliptical galaxies the spread is reduced by perhaps 20-30 percent (stretch factors ~ 0.7 to 1.050 [15]. (2) The mean luminosities of SNe Ia observed in spiral galaxies are clearly higher than those of elliptical galaxies [15,16]. A very significant factor here is the absence of the brightest SNe Ia in elliptical and S0 galaxies. While this would appear to be a population based correlation having to do with e.g. the metallicity or the age of the underlying stellar component, the cause of this is not unambiguous. It is the factors that can impact the diversity – as reflected in the variations in peak luminosities – which we wish now to identify and discuss.

## 3. Creating Diversity

A critical clue to understanding the observed diversity of Type Ia is the recognition that their peak luminosities are proportional to the mass of $^{56}$Ni ejected [11]. For $^{56}$Ni to be a dominant feature of supernova ejecta itself imposes constraints on supernova models: (1) the peak temperature must be sufficient (typically $T > 4\text{-}5\text{x}10^9$ K) to insure that nuclear transformations proceed rapidly enough to burn the available carbon and oxygen fuels to iron-peak nuclei on a hydrodynamic timescale; (2) the densities must be high enough to insure that these high temperatures can be achieved as a consequence of the advance of the burning front through the outer regions of the star (typically, for densities below ~ $10^6$ g cm$^{-3}$, the nuclear burning products can involve rather intermediate mass nuclei in the range silicon to calcium or titanium); and (3) the degree of neutron enrichment of the matter ($Y_e$) must lie in the range compatible with the synthesis of $^{56}$Ni and not more neutron-rich elements like $^{54}$Fe and $^{58}$Ni, or even the direct production of $^{56}$Fe itself (this constrains the densities in the critical regions to values below ~ 3 x $10^9$ g cm$^{-1}$, so that weak interactions will not neutronize too large a fraction of the core mass).

Factors that can be understood to be capable of influencing the light curves then include anything that can impact $^{56}$Ni production. Two important considerations are the following:

(1) The primordial composition of the white dwarf star can introduce a level of neutronization (in the form of $^{22}$Ne) over the entire inner regions of the white dwarf that will reduce the $^{56}$Ni abundance at the expense of forming greater concentrations of $^{54}$Fe and $^{58}$Ni (Timmes et al. [18]). This neutron enrichment arises from first the conversion of CNO isotopes to $^{14}$N during the hydrogen burning phase and second the conversion of this to $^{22}$Ne by the reactions

$$^{14}\mathrm{N}(\alpha,\gamma)^{18}\mathrm{F}(e^{+}\nu)^{18}\mathrm{O}(\alpha,\gamma)\ ^{22}\mathrm{Ne}$$

For matter of solar composition, this yields a $^{22}$Ne concentration of about 2.5 % by mass.

(2) The density at which nuclear burning proceeds in the outermost regions also constrains $^{56}$Ni production. For densities < $10^6$ g cm$^{-1}$, the peak temperature is not sufficient to burn to nickel, and the main burning products will typically be intermediate mass nuclei (~silicon-to calcium) at the expense of $^{56}$Ni, and dim Type Ia supernovae will result.

The manner in which these factors can influence peak supernova brightness is evident from some features of Figure 2 below, where the abundance pattern is that predicted for the "carbon deflagration" nucleosynthesis calculations of Thielemann et al. [19]. For this model, the velocities achieved in the outermost layers containing intermediate mass nuclei reached ~ 15,000 km s$^{-1}$. Note the high degree of neutron enrichment of the innermost layers ( ~ 0.2 solar masses) by means of electron captures, the broad region of mass ( ~0.8 solar masses) dominated by $^{56}$Ni, and the ~ 0.4 solar masses of intermediate mass nuclei in the outermost regions.

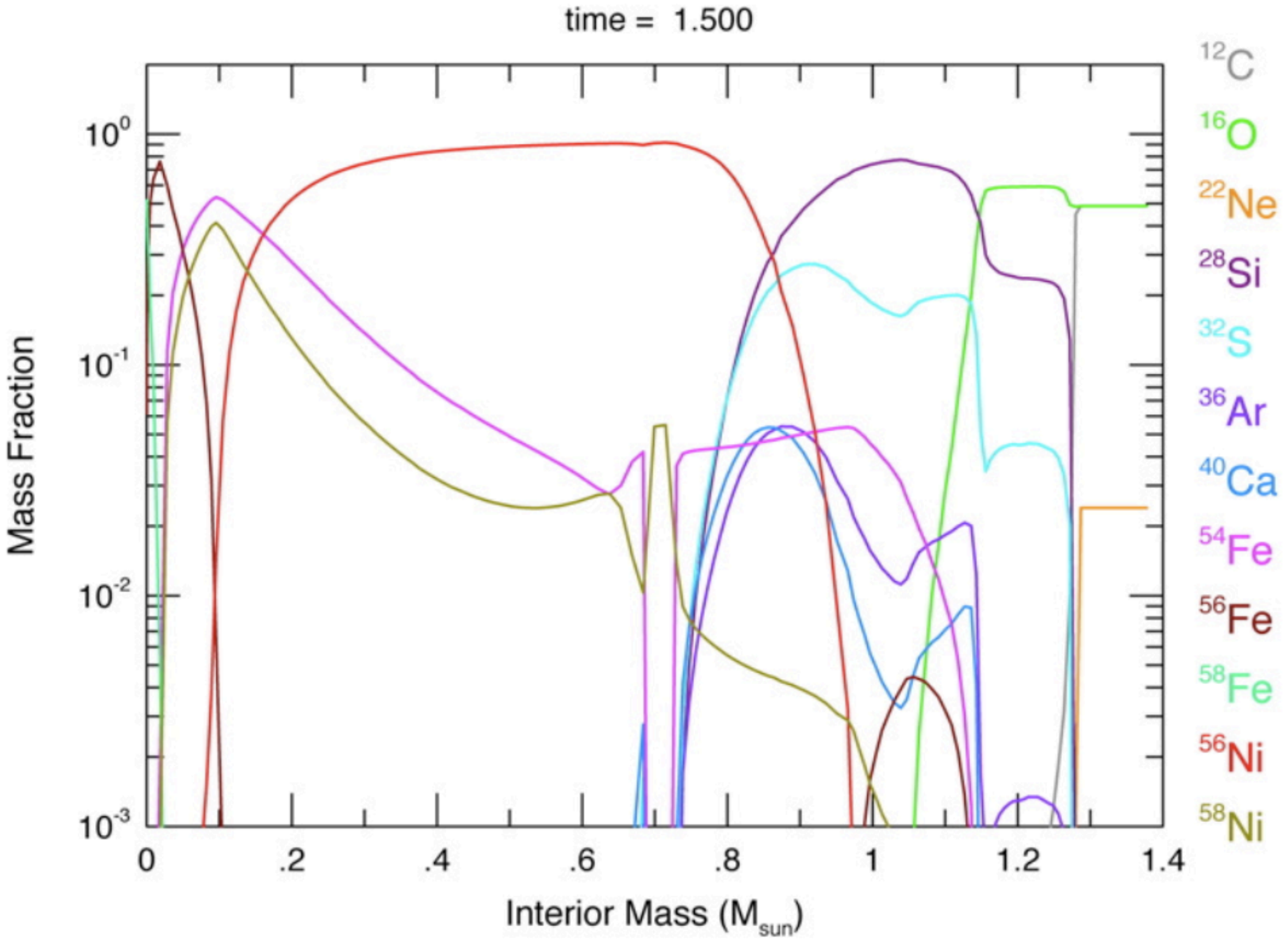

Figure 2: The composition of the ejected matter predicted for the carbon-deflagration model of Thielemann, Nomoto, and Yokoi [19]. The velocity history of the ejecta began with low velocities that increased systematically as the burning front continued outward through the star, such that the velocities of the outermost ejecta reached values ~ 15,000 km s$^{-1}$, consistent with observations. (Figure courtesy F.-X. Timmes)

**4, Implied Constraints on the Diversity of Interesting Type Ia Supernova Models:**

The history of nucleosynthesis models has been strongly guided and constrained by observations. The early 'carbon detonation' model of Arnett [20] was successful in predicting an iron peak *in situ* that was dominated by $^{56}$Ni and for which many isotopes in this region were synthesized in approximately solar proportons [21], but failed in that virtually all the mass was ejected as $^{56}$Ni. This was in conflict with observations by Branch [ 22] of the presence of "a mixture of intermediate mass elements from carbon to calcium" in the ejecta.

A significant and needed characteristic of 'deflagration' models is the (pre-)expansion of the outermost regions of the white dwarf prior to the arrival of the burning front. It is this feature that leads to the synthesis of intermediate mass nuclei, in that the temperatures achieved in the burning regions at lower densities are insufficient to drive burning to iron. The high velocities of the early ejecta confirm, however, that the velocities in the outer regions must approach sound speed. This has led to considerations of models in which the flame begins as a deflagration and undergoes a "deflagration to detonation transition" or DDT [23, 24, 25].

Another interesting and studied model for Type Ia supernovae is the 'gravitationally confined detonation' or GCD model [26, 27]. In this mechanism, ignition occurs at one or several off-center points, resulting in a burning bubble of hot gas that is propelled rapidly to the surface by buoyancy, breaks through the surface of the star, and collides at a point opposite the breakout on the stellar surface. Detonation conditions are then achieved as a result of the inwardly directed jet that is produced by the compression of unburnt surface matter when the surface flow collides with itself. A diversity reflected here in the range of peak brightness can naturally arise as a consequence of the number of ignition points and the positioning of these ignition points. Pre-expansion here again will yield lower density conditions in the outermost regions, constraining nickel production and producing concentrations of intermediate mass nuclei

All of the models we have discussed above are 'single degenerate' models that are generally assumed to involve a Chandrasekhar mass carbon-oxygen white dwarf in a close binary system. A challenging question in all cases is to identify the nature of the progenitor system within which the white dwarf has grown to the critical mass, presumably by accretion from its binary companon. No definitive model for such growth has yet to have been identified. For this reason, there is renewed interest in "sub-Chandrasekhar" mass white dwarf models that involve either off-center ignition of helium at sub-Chandrasekhar masses or perhaps rather white dwarf-white dwarf collosions/mergers in double degenerate systems. As examples, we note the recent studies by Fink et al. [28] and Sim et al. [29] of detonations/double-detonation in sub-Chandrasekhar CO white dwarfs. The diversity in these models can result as well simply from the variations in white dwarf mass.

In conclusion, we emphasize again the important feature of SNe Ia outbursts – that it is the energy release in the decay of $^{56}$Ni that is reflected in the observed peak luminosities of these events - and that this provides an extremely important tool to explore and to constrain theoretical models of these events.

**Acknowledgments:**
This work is supported in part at the University of Chicago by the National Science Foundation under Grant PHY 02-16783 for the Frontier Center "Joint Institute for Nuclear Astrophysics" (JINA), and in part at the Argonne National Laboratory by the U.S. Department of Energy, Office of Nuclear Physics, under contract DE-AC02-06CH11357.